\documentclass[epj]{svjour}
\usepackage{mathrsfs}
\usepackage{graphicx}
\usepackage{amsfonts}
\usepackage{amssymb}%
\providecommand{\Journal}[4] {#1 {\bf#2}, #4 (#3)}
\providecommand{\PLB}{Phys. Lett. B} %
\providecommand{\PRL}{Phys. Rev. Lett.} %
\providecommand{\PRD}{Phys. Rev. D}
\providecommand{\PIC}{Physics in Canada}
\providecommand{\NT}{Nature}
\providecommand{\NPB}{Nucl. Phys. B} %
\providecommand{\EPJC}{Eur. Phys. J. C} %
\providecommand{\Jh} {JHEP}%
\providecommand{\IB}{ibid}%
\providecommand{\IJMP}{Int. J. Mod. Phys.}%
\providecommand{\LRR}{Living Rev. Rel.}%

\newcommand{\be}{\begin{equation}}
\newcommand{\ee}{\end{equation}}
\newcommand{\bea}{\begin{eqnarray}}
\newcommand{\eea}{\end{eqnarray}}

\newcommand{\hf}{\frac{1}{2}}
\newcommand{\al}{\alpha}
\newcommand{\e}{\epsilon}
\newcommand{\nn}{\nonumber\\}

\newcommand{\pt}{\partial}

\begin{document}
\title{Eikonal equation of the Lorentz-violating Maxwell theory}
\author{Zhi Xiao\inst{1}, Lijing Shao\inst{1} \and Bo-Qiang Ma\inst{1,2,}
\thanks{\emph{Electronic address:} mabq@pku.edu.cn}%
}                         
%
%
\institute{School of Physics and State Key Laboratory of Nuclear
Physics and Technology, Peking University, Beijing 100871, China
\and Center for High Energy Physics, Peking University, Beijing
100871, China}
\date{Received: date / Revised version: date}
%
\abstract{ We derive the eikonal equation of light wavefront in the
presence of Lorentz invariance violation (LIV) from the photon
sector of the standard model extension (SME). The results obtained
from the equations of $\mathbf{E}$ and $\mathbf{B}$ fields
respectively are the same. This guarantees the self-consistency of
our derivation. We adopt a simple case with only one non-zero LIV
parameter as an illustration, from which we find two points. One is
that, in analogy with Hamilton-Jacobi equation, from the eikonal
equation, we can derive dispersion relations which are compatible
with results obtained from other approaches. The other is that, the
wavefront velocity is the same as the group velocity, as well as the
energy flow velocity. If further we define the signal velocity $v_s$
as the front velocity, there always exists a mode with $v_s>1$,
hence causality is violated classically. 
Thus our method might be useful in the analysis of Lorentz violation in
QED in terms of classical causality .
} 
\maketitle

\onecolumn

\section{Introduction\label{sec:1}}

The constant speed of light in vacuum is one of the basic
assumptions of special relativity, and it serves as a crucial
ingredient of Lorentz symmetry, which is a cornerstone of modern
physics. However, currently there is a revival of interest in the
possibility of varying speed of light, or in other words, a tiny
deviation from exact Lorentz invariance, in the search of
so-called quantum gravitational phenomena.

As the typical quantum gravitational scale, {\it i.e.}, the Planck
scale $E_\mathrm{Planck}=\sqrt{c\hbar/G}c^2$, is practically
unattainable from conventional accelerator experiments, people turn
to search for phenomenologically accessible effects from quantum
gravity~\cite{QGPe} at relatively low energy scales. One of them is the possibility of Lorentz
invariance violation (LIV). The possibility that quantum gravity may
leave a tiny imprint of LIV at relatively low energies was observed
by many authors from various approaches to quantum gravity. These
include string field theory, where the tachyon field may induce an
instability to the naive Lorentz invariant vacuum and translate it
into the potential of a tensor field. As a consequence, the tensor field
acquires a vacuum expectation value~\cite{string} and breaks Lorentz
invariance. Later, by incorporating various LIV coefficients into
the standard model, the theory was developed into an effective field
theory (EFT), called standard model extension (SME)~\cite{SME}.
Other approaches include spin network calculation in loop
gravity~\cite{loop}, deformed special relativity~\cite{DSR}, foamy
structure of space-time~\cite{foam},
noncommutative field theory~\cite{Nonc}, emergent 
gravity~\cite{emer}, and the recently suggested Horava-Lifschitz
gravity~\cite{Horava}. All of them suggest that tiny LIV may be a
signature of new physics.

Fortunately, LIV not only resides in 
theoretical considerations, it also becomes experimentally testable,
and has already been tested to very high accuracy in various sectors
of the standard model (for a review, see Ref.~\cite{Data}). From an
experimental viewpoint, people have already been able to severely
constrain the linear order LIV correction to the photon group
velocity, which is suppressed by the ratio of experimental energy
$E$ to a large mass scale, 
{\it e.g.},
$E/M_\mathrm{Planck}$ (see {\it e.g.} Refs.~\cite{FLAT,mine,LZB,shaoma,DUHECR}
). 

From the EFT viewpoint,
conventionally non-renormalizable operators can be naturally
suppressed by a large mass scale, 
where new physics might come in. So the mentioned tiny linear
correction may come from dimension-5 operators~\cite{EFT}. However,
if one assumes CPT symmetry and the anisotropic scaling between
space and time~\cite{Horava} (also termed weighted power
counting~\cite{weight}), or alternatively imposes CPT invariance and
the supersymmetric constraints~\cite{susy}, one can naturally expect
that the leading order correction comes from dimension-6 operators,
thus can evade current experimental constraints~\cite{Earbi}. On the other side, there is no clear
theoretical reason to pin down the values of dimension-3 and -4 LIV
operators, though they have already been extensively studied and
found to suffer severe constraints from
experiments~\cite{Data,CSME}. Hence it is still valuable to consider
them both theoretically and experimentally, and it inspires us to
revisit the renormalizable LIV operators~\cite{SME}. In this paper,
we mainly focus on the CPT-even part of the photon sector in the
framework of SME.

The paper is organized as follows. In section 2, we give a brief
review on the renormalizable photon sector of SME. With some {\it ad
hoc} assumptions for simplification, we derive the modified Maxwell
equations as our starting point. In section 3, following the
derivation provided by Fock~\cite{Fock}, we obtain the eikonal
equation of the modified Maxwell equations implicitly. In section 4,
by adopting a much simpler case with only one non-zero LIV parameter
as an illustration, we give the eikonal equation explicitly. From
the derivation, we find it is self-consistent, because 
one can obtain the same result from 
equations relevant to either an $\textbf{E}$ or $\textbf{B}$ field, respectively.
It is also consistent with other approaches, as the
dispersion relations obtained from our procedure and those from
others are the same. Meanwhile, in this simple case, we find that
the group velocity equals the wavefront velocity, as well as the
energy flow velocity. As the derivation of eikonal equation is
obviously not restricted to the simple model, 
the method may be useful in the classical causality analysis in
LIV extension of quantum electrodynamics. Section 5 summarizes the
results obtained in this paper and gives a further discussion on LIV
topics.

\section{A brief review of the photon part of SME\label{sec:2}}

SME, including in principle all possible LIV terms in the
framework of EFT~\cite{SME}, is a well-motivated testable approach
to LIV physics, and it has been extensively tested in various
sectors of the standard model~\cite{Data}. Within SME, the
renormalizable photon sector is one of the most tested
parts~\cite{CSME,pbound}. Its Lagrangian reads

\bea\label{rp}&&
    \mathcal{L}_\mathrm{photon}=  
    -\frac{1}{4}F_{\mu\nu}F^{\mu\nu}-\frac{1}{4}(k_F)_{\kappa\lambda\mu\nu}F^{\kappa\lambda}
         F^{\mu\nu}+\frac{1}{2}(k_{AF})_\kappa\epsilon^{\kappa\lambda\mu\nu}A_\lambda
         F_{\mu\nu},
\eea
 where the coefficients $k_F$ and $k_{AF}$ characterize the
violation of Lorentz symmetry. From Eq.~(\ref{rp}), the equation of
motion is deduced as \bea\label{eom}
    &&
  \partial^\alpha
    F_{\mu\alpha}+(k_F)_{\mu\alpha\beta\gamma}\partial^\alpha
    F^{\beta\gamma}+(k_{AF})^\alpha\epsilon_{\mu\alpha\beta\gamma}F^{\beta\gamma}=0.
\eea

With the 3+1 decomposition, we define
 \bea\label{prde} && (k_{DE})^{jk}\equiv-2(k_F)^{0j0k},\quad
         (k_{HB})^{il}\equiv\hf(k_F)^{jkmn}\e^{ijk}\e^{lmn}, \nn &&
         (k_{DB})^{jk}\equiv-(k_{HE})^{kj}\equiv\hf(k_F)^{0jmn}\e^{kmn},
         \eea
where the Latin indices run over the three spatial coordinates, from
1 to 3. After subtracting the trace part of the first two matrices,
we define further the following traceless matrices which are
frequently used in this paper, \bea\label{myde} &&
{\beta_E}^{ij}=(k_{DE})^{ij}-\alpha\delta^{ij},\quad
  {\beta_B}^{ij}=(k_{HB})^{ij}+\alpha\delta^{ij}, \quad
  \gamma^{ij}= (k_{DB})^{ij}=-(k_{HE})^{ij},
         \eea
where \be\label{trace}
\alpha=\frac{1}{3}\mathrm{tr}(k_{DE})=-\frac{1}{3}\mathrm{tr}(k_{HB}).\ee
Note that the $(k_F)_{\mu\alpha\beta\gamma}$ tensor has the same
symmetry as the Riemann tensor,
$R_{\mu\alpha\beta\gamma}$~\cite{SME,LVE}. Due to the double
tracelessness of $k_F$ and the Bianchi identities
$(k_F)_{\mu[\nu\rho\sigma]}=0$, we can get Eq.~(\ref{trace}) and
$\mathrm{tr}[\gamma]=0$, respectively. The first two matrices in
Eq.~(\ref{myde}) can be easily shown as symmetric, {\it i.e.},
$({\beta_E})^T={\beta_E}$ and $({\beta_B})^T={\beta_B}$. Since the
$(k_{AF})_\mu$ term may cause theoretical instabilities~\cite{SME}
and has suffered severe astrophysical constraints~\cite{CFJ}, we
abandon it from now on. Moreover, as $\gamma^{ij}$ mixes ${\bf E}$
and ${\bf B}$ fields and may cause further complications in
calculation, we simply set $\gamma^{ij}=0$. Further, we also set
$\al=0$ as an \emph{ad hoc} assumption. Under these simplifications,
there are only 10 parameters left, residing in the traceless
symmetric matrices ${\beta_E}$ and ${\beta_B}$. Finally, we can
write down the Lagrangian in terms of {\bf E} and {\bf B} as
\bea\label{SL}
\mathcal{L}_\mathrm{photon}=\frac{1}{2}(\vec{E}^2-\vec{B}^2)+\hf\left((\beta_E)^{jk}E^jE^k-(\beta_B)^{jk}B^jB^k\right),
\eea where the electric field {\bf E} and the magnetic field {\bf B}
are defined conventionally as %
\bea\label{FEB}
B^i=-\hf\epsilon^{ijk}F_{jk}, \quad E^i=-{\partial_t}A^i-
{\triangledown_i} \phi = F^{i0} .\eea

From Eq.~(\ref{SL}), we can deduce the modified Maxwell equations,
\bea\label{MME1}&&
\triangledown\cdot\bf{E}+\partial_i({\beta_E})^{ij}\bf{E}^j=0,
~~~~\quad
-\bf{\dot{E}}^i-({\beta_E})^{ij}\bf{\dot{E}}^j+\epsilon_{ijk}\partial_j\bf{B}^k+\epsilon_{ijk}\partial_j({\beta_B})^{kl}\bf{B}^l=0,
\eea
and the equations
\bea\label{MME2}&&
\triangledown\cdot\bf{B}=0,~~\quad ~~~
\triangledown\times\bf{E}+\bf\dot{B}=0
\eea
come from the 3+1 decomposition of Bianchi
identity $\partial_\mu{^*F}^{\mu\nu}=0$, where
${^*F}^{\mu\nu}=\hf\epsilon^{\mu\nu\beta\gamma}F_{\beta\gamma}$. The
above four equations are our starting points of derivation in the
next section.

Note that Eq.~(\ref{MME1}) has the analogy of electrodynamics in a
homogeneous anisotropic media~\cite{SME} if we regard
$(\beta_B)^{ij}$ as the inverse of magnetic permeability and
$(\beta_E)^{ij}$ as the dielectric constant, {\it i.e.}, as defined
in Ref.~\cite{LVE},
$\bf{D}^i=(\delta^{ij}+{\beta_E}^{ij})\bf{E}^j,\quad
\bf{H}^i=(\delta^{ij}+{\beta_B}^{ij})\bf{B}^j$. From this analogy,
we expect that the differential equation satisfied by the light
wavefront in LIV vacuo should have a similar form as the eikonal
equation in an anisotropic medium. Actually, we will see that
this is indeed the case in section 4.

Now we turn to the discussion of the energy-momentum tensor, which
is useful for the discussion of causality in section 4. The
conventional procedure to obtain a symmetric energy-momentum tensor
is through the so-called Belinfante tensor~\cite{Weinberg}
\bea\label{symEM}
\Theta^{\mu\nu}=T^{\mu\nu}-i\hf\partial_\kappa[\frac{\partial\mathcal
{L}}{\partial(\partial_\kappa\Psi_l)}(\mathscr{J}^{\mu\nu})_l^{~m}\Psi_m-\frac{\partial\mathcal
{L}}{\partial(\partial_\mu\Psi_l)}(\mathscr{J}^{\kappa\nu})_l^{~m}\Psi_m-\frac{\partial\mathcal
{L}}{\partial(\partial_\nu\Psi_l)}(\mathscr{J}^{\kappa\mu})_l^{~m}\Psi_m],
\eea where \bea T^{\mu\nu}=\frac{\partial\mathcal
{L}}{\partial(\partial_\mu\Psi_l)}\partial^\nu\Psi_l-\eta^{\mu\nu}\mathcal{L}.
\eea However, this does not work in the presence of LIV, as
pointed out by Colladay and Kostelecky~\cite{SME}.

In fact, from Eq.~(\ref{symEM}), we can get the energy-momentum
tensor corresponding to the Lagrangian in Eq.~(\ref{rp}) without the
CPT odd term, 
\bea\label{EMEM}&&
\Theta^{\mu\nu}=-[F^{\mu\alpha}F^\nu_{~\alpha}+(k_F)^{\alpha\beta\mu\delta}{F^\nu_{~\delta}}F_{\alpha\beta}]-\eta^{\mu\nu}\mathcal{L}\nn
&&
~~~~~=-[F^{\mu\alpha}F^\nu_{~\alpha}+(k_F)^{\alpha\beta\mu\delta}{F^\nu_{~\delta}}F_{\alpha\beta}]+
\frac{1}{4}\eta^{\mu\nu}[(k_F)_{\alpha\beta\gamma\delta}F^{\alpha\beta}F^{\gamma\delta}+F^{\alpha\beta}F_{\alpha\beta}].
\eea In the above derivation, we have used the matrix representation of the homogeneous
Lorentz algebra for a covariant vector field, \be
(\mathscr{J}^{\mu\nu})_\rho^{~\sigma}=i(\delta^\mu_{~\rho}\eta^{\nu\sigma}-\delta^\nu_{~\rho}\eta^{\mu\sigma}).
\ee

We find that 
due to the presence of the second asymmetric 
term,
$-(k_F)^{\alpha\beta\mu\delta}{F^\nu_{~\delta}}F_{\alpha\beta}$, one
can no longer obtain a symmetric energy-momentum tensor. This may be
one feature of LIV theories. One consequence is that the definition
of the conserved 4-momentum density is $\Theta^{0\mu}$, instead of
$\Theta^{\mu0}$. Actually, with direct calculation it is easily
checked that $\pt_\rho\Theta^{\rho\mu}=0$, hence
$d(\int{d^3x\Theta^{0\mu}})/dt=0$ (for details, see Appendix A). In
contrast, this is not valid for $\Theta^{\mu0}$. The spatial part
of the latter 
is defined as the generalized Poynting vector~\cite{SME}. For the
convenience of our discussions in section 4, we give those
components of the energy-momentum tensor explicitly in terms of {\bf
E}, {\bf B} fields, 
\bea\label{EnDe}&&
\Theta^{00}=\hf({E}^2+{B}^2)-(k_F)^{0j0k}E^jE^k+\frac{1}{4}(k_F)^{ijkl}\epsilon^{ijm}\epsilon^{kln}B^mB^n\nn
&&~~~~~=\hf({E}^2+{B}^2+E^j(\beta_E)^{jk}E^k+B^m(\beta_B)^{mn}B^n),\\
\label{MoDe}&&
\Theta^{0i}=-F^{0j}F^i_{~j}-2(k_F)^{0j0k}F^i{~j}F_{0k}-(k_F)^{0jkl}F^i_{~j}F_{kl}\nn
&&~~~~~=\epsilon^{ijk}E^jB^k+\epsilon^{ijk}(\beta_E)^{jl}E^lB^k,\\
\label{Poyn}&&
\Theta^{i0}=-F^{ij}F^0_{~j}-(k_F)^{ijkl}F^0_{~j}F_{kl}-2(k_F)^{ij0l}F^0_{~j}F_{0l}\nn
&&~~~~~=\epsilon^{ijk}E^jB^k+\epsilon^{ijk}E^j(\beta_B)^{kl}B^l,
\eea
where we have used the reparametrization of Eqs.~(\ref{prde})
and (\ref{myde}), and the simplifications corresponding to the
Lagrangian in Eq.~(\ref{SL}). From Eqs.~(\ref{MoDe}) and
(\ref{Poyn}), it is apparent that
due to the loss of Lorentz invariance, characterized by $\beta_E$
and $\beta_B$, generally the energy-momentum tensor is not symmetric, {\it
i.e.}, $\Theta^{\mu\nu}-\Theta^{\nu\mu}=[
(k_F)^{\alpha\beta\mu\delta}{F^\nu_{~\delta}}-(k_F)^{\alpha
\beta\nu\delta}{F^\mu_{~\delta}}]F_{\alpha\beta}\neq0$.

It has been well known that, in the Lorentz invariant theory, the
energy-momentum tensor should be symmetric due to the following
reasons~\cite{Ryder}:
\begin{enumerate}
\renewcommand{\labelenumi}{(\theenumi)}
\item
When it couples with gravity in curved space-time, the
symmetrical Einstein tensor $G_{\mu\nu}$ and metric tensor
$g_{\mu\nu}$ automatically imply a symmetric $\Theta^{\mu\nu}$ due
to the Einstein equation;
\item
The current conservation of
$M^{\rho\mu\nu}=\hf(x^\nu\Theta^{\rho\mu}-x^\mu\Theta^{\rho\nu})$ of
the angular momentum tensor $\Sigma^{\mu\nu}=\int{d^3x}M^{0\mu\nu}$
also requires a symmetric $\Theta^{\mu\nu}$. \end{enumerate}

However, in our case, LIV undermines these two reasons;
\begin{enumerate}
\renewcommand{\labelenumi}{(\theenumi)}
\item
It is necessary to take LIV effects into account in gravity side for
a consistent theory, however in this case, it is the local Lorentz
invariance involved. Moreover, when coupled with gravity, LIV
parameters should be promoted to dynamical fields instead of being
constants~\cite{MTLI}, and the compatible framework to incorporate
LIV in general relativity is the Riemann-Cartan
geometry~\cite{RGLI}, where the extended Einstein tensor is no
longer symmetric. The antisymmetric part of the Einstein tensor and
the energy-momentum tensor is consistently linked through field
equations~\cite{RGLI}. Thus no inconsistency arises when an
asymmetric $\Theta^{\mu\nu}$ couples with gravity in a spontaneous
LIV theory~\cite{RGLI}.
\item
Due to the breaking of Lorentz invariance, the generator of the
Lorentz group, {\it i.e.}, the angular momentum tensor, is no longer
conserved in general, thus no constraints are imposed here. If
one assumes that only boost invariance is broken, {\it i.e.},
rotational invariance is still preserved. It means that only $\al$
and $k_{AF}^0$ are non-zero in Eq.~(\ref{rp}). Then in this case,
one can prove $\Theta^{ij}=\Theta^{ji}$, which is required by the
conservation of the spatial part of the angular momentum tensor,
{\it i.e.}, $d\Sigma^{ij}/dt=0$. However, in the presence of LIV, it is
only valid in a particular reference frames, {\it e.g.}, CMB reference, as the
transformation from one inertial frame, in which the LIV photon equation is isotropic,
cannot remain isotropic in other inertial frames~\cite{update}.
Here we present the proof in Appendix A.
\end{enumerate}

\section{A derivation of eikonal equation\label{sec:3}}

In order to derive the eikonal equation of the modified
electrodynamics, it is necessary to review how this is achieved in
Lorentz invariant Maxwell theory. We follow the derivation of
Fock~\cite{Fock}. The basic principle is that the fields on the
wavefront, which is a space-like 2-dimension surface, must be
noncontinuous in the normal direction of the surface. The reason is
that the surface is an interface of two regions, one already
influenced by the field perturbation and the other remaining
intact. Thus the time derivatives of the fields on the wavefront
should be singular.

Taking a right-moving plane wave as an example, the wavefront is
an infinite plane perpendicular to its constant wavevector $\vec{k}$
at any instant. The left region of the plane is a region with
non-vanishing electromagnetic fields, while the right side is a
region with vanishing fields.
This conveys the fact that light propagates with a finite speed,
thus at any instant it cannot propagate further to the right of the wavefront plane.
Since an electromagnetic field evolves dynamically, it is more helpful to
analyze from a 4-dimensional point of view, rather than from the view at
some fixed instant in 3-spatial dimensions.

For any instant $t$, the wavefront in the 3-spatial dimension can be
written as \bea\label{exlic} t=f(x,y,z). \eea This can also be
viewed in a dynamical way as an evolving 2-dimensional space-like
surface $f(x,y,z)-t=0$. When we write Eq.~(\ref{exlic}) in an implicit form,
\bea\label{imlic} w(t;x,y,z)=0, \eea
the wavefront becomes a slice of 3-dimensional hyper-surface in flat
space-time in the 4-dimensional point of view. So the 2-dimensional space-like
wavefront is just a time slice of the null-surface (light-cone),
and the non-continuity of fields on the wavefront is nothing more
than the (classical) causality requirement. To be more specific, the
fields inside the light-cone must be continuous since they are
causally connected. Using Maxwell equations, one can obtain from the
fields on any given surface in 3-spatial dimensions inside the light-cone,
the fields on an infinitesimally close surface.
However, this cannot be done for fields lying on the surface
outside the light-cone. So the instant time slice of the 
hyper-surface $w(t;x,y,z)=0$ 
becomes an interface between causally connected and non-connected
regions, hence the derivatives of fields on the interface become
singular. We use the above principle to derive the so-called eikonal
equation with the same logic as that in the Lorentz invariant theory.

The Lorentz invariant eikonal equation reads \bea\label{conv}
(\triangledown{f(x,y,z)})^2=n^2, \eea  where in vacuum, $n$ equals
1 in our unit ({\it i.e.} $c=1$). In a more explicit Lorentz
invariant form, it reads \bea\label{LILC}
(\frac{\pt{w}}{\pt{t}})^2-(\triangledown{w})^2=0. \eea We show in
Appendix B how to get Eq.~(\ref{LILC}) from Eq.~(\ref{conv}).

Now we turn to derive eikonal equation in the presence of LIV. Any
field $u(x,y,z,t)$ on a 3-dimensional hyper-surface in the
4-dimensional Minkowski space is \be\label{arbf}
u(t,x,y,z)=u(f(x,y,z);x,y,z)\equiv u_0(x,y,z). \ee Therefore, we
have \be
\frac{\pt{u_0}}{\pt{x^i}}=\frac{\pt{u}}{\pt{x^i}}+\frac{\pt{u}}{\pt{t}}\frac{\pt{f}}{\pt{x^i}},
\ee where $x^i$ with Latin index $i$ running over 1 to 3, represents $x,~y,~z$
respectively. Thus the spatial derivatives of the electric and
magnetic fields on the hyper-surface are \bea\label{EB}
\frac{\pt\bf{E}_0^i}{\pt{x^j}}=\frac{\pt\bf{E}^i}{\pt{x^j}}+\pt_jf\bf{\dot{E}}^i,~~~
\quad
\frac{\pt\bf{B}_0^i}{\pt{x^j}}=\frac{\pt\bf{B}^i}{\pt{x^j}}+\pt_jf\bf{\dot{B}}^i.
\eea

From the linearity of
Eq.~(\ref{EB}), we have \bea\label{LCE1}
\triangledown\times\bf{E}_0=\triangledown\times\bf{E}+\triangledown{f}\times\bf{\dot{E}}.
\eea Then by using the second equation of the Bianchi identity (\ref{MME2}), we get
\bea\label{LCE2}
\triangledown\times\bf{E}_0=-\bf{\dot{B}}+\triangledown{f}\times\bf{\dot{E}}.
\eea Similarly, one can obtain from Maxwell Eqs.~(\ref{MME1}) and
(\ref{MME2}) the set of equations below
\bea\label{LCE2a}&&
\triangledown\cdot((1+\beta_E)\cdot\bf{E}_0)=\triangledown{f}\cdot((1+\beta_E)\cdot\bf{\dot
E}),\quad  ~~~~
\triangledown\times\bf{E}_0=-\bf{\dot{B}}+\triangledown{f}\times\bf{\dot{E}};
\\
\label{LCE2b}&&
\triangledown\cdot\bf{B}_0=\triangledown{f}\cdot\bf{\dot  B},\quad
~~~~~~~~
\triangledown\times((1+\beta_B)\cdot\bf{B}_0)=((1+\beta_E)\cdot\bf{\dot{E}})+\triangledown{f}\times((1+\beta_B)\cdot\bf{\dot{B}}).
\eea
Then by multiplying the second equations of (\ref{LCE2a}) and
(\ref{LCE2b}) respectively with $\triangledown{f}$, we have
\bea\label{LCE3a}&&
\triangledown{f}\cdot\triangledown\times\bf{E}_0=-\triangledown{f}\cdot\bf{\dot{B}},~~~~~~~~~~~~~~~~~~\\
\label{LCE3b}&&
\triangledown{f}\cdot\triangledown\times((1+\beta_B)\cdot\bf{B}_0)=\triangledown{f}\cdot((1+\beta_E)\cdot\bf{\dot{E}}).
\eea
Comparing the equations above with the first equations of
(\ref{LCE2a}) and (\ref{LCE2b}), we find the following relations
\bea\label{LCE4a}&&
\triangledown{f}\cdot\triangledown\times\bf{E}_0+\triangledown\cdot\bf{B}_0=0,~~~~~~~~~~~~~~~~~~~~~\\
\label{LCE4b}&&
\triangledown{f}\cdot\triangledown\times((1+\beta_B)\cdot\bf{B}_0-\triangledown\cdot((1+\beta_E)\cdot\bf{E}_0)=0.
\eea

When $f(x,y,z)$ is a constant, {\it i.e.}, by specifying a
particular
instant, 
Eqs.~(\ref{LCE4a}) and (\ref{LCE4b}) become the first equations of
the original field equations (\ref{MME1}) and (\ref{MME2})
respectively. This means that we need to specify proper initial
conditions.

In order to extract more information, we multiply the curl of $E_0$
by the tensor $(1+\beta_B)$, then calculate its cross product with
$\triangledown{f}$. By utilizing the second equations of
(\ref{LCE2a}) and (\ref{LCE2b}) successively, we get
\bea&&
\triangledown{f}\times[(1+\beta_B)\cdot\triangledown\times\bf{E}_0]=\triangledown{f}\times[(1+\beta_B)\cdot(\triangledown{f}\times\bf{\dot{E}}-\bf{\dot{B}})]\nn
&&=(1+\beta_E)\cdot\bf{\dot{E}}-\triangledown\times[(1+\beta_B)\cdot\bf{B}_0]+\triangledown{f}\times[(1+\beta_B)\cdot(\triangledown{f}\times\bf{\dot{E}})].
\eea
After arranging it in such a way that all time derivatives of
{\bf E} are on one side, and $E_0$ and $B_0$ on the other side,
we have
\bea\label{Eligh}
(1+\beta_E)\cdot\bf{\dot{E}}+\triangledown{f}\times[(1+\beta_B)\cdot(\triangledown{f}\times\bf{\dot{E}})]=\triangledown{f}\times[(1+\beta_B)\cdot\triangledown\times\bf{E}_0]
+\triangledown\times[(1+\beta_B)\cdot\bf{B}_0].
\eea

Now it is apparent that the right-hand side of (\ref{Eligh})
includes the wavefront function $f$ and the electromagnetic fields $E_0$ and $B_0$ with values on the 
wavefront, while the left-hand side includes only $\bf{\dot{E}}$ and $f$.
With given wavefront function $f$ and the fields $E_0$ and $B_0$,  
$\bf{\dot{E}}$ on the interface must be singular, as we have already argued. Otherwise one can
obtain the fields on an infinitesimal close surface outside
the light-cone from field equations, which obviously violates the
causality requirements. Thus, if we view the left-hand side of
Eq.~($\ref{Eligh}$) as a matrix equation, \bea\label{MaE}
M_e\cdot\bf{\dot{E}}\equiv(1+\beta_E)\cdot\bf{\dot{E}}+\triangledown{f}\times[(1+\beta_B)\cdot(\triangledown{f}\times\bf{\dot{E}})],
\eea where $M_e$ is a tensor/matrix, 
then the determinant of $M_e$ must be zero, {\it i.e.},
$\mathrm{Det}(M_e)=0$. Otherwise, in principle, one would get a
non-singular $\bf{\dot{E}}$ by solving Eq.~(\ref{Eligh}).

Similarly, we can obtain the corresponding equation of the {\bf B}
field, \bea\label{Bligh}
\bf{\dot{B}}+\triangledown{f}\times[(1+\beta_E)^{-1}\cdot\{\triangledown{f}\times\left((1+\beta_B)\cdot\bf{\dot{B}}\right)\}]=
\triangledown{f}\times\{(1+\beta_E)^{-1}\cdot\left(\triangledown{f}\times[(1+\beta_B)\cdot\bf{B}_0]\right)\}-\triangledown\times\bf{E}_0,
\eea where we have presumed the existence of the inverse of
$(1+\beta_E)$, due to the fact that all of the 10 parameters, $(\beta_E)^{ij}$ and
$(\beta_B)^{ij}$, must be very tiny compared to 1 to meet the
stringent experimental constraints.

One can also define \bea\label{MaB}
M_b\cdot\bf{\dot{B}}\equiv\bf{\dot{B}}+\triangledown{f}\times
\left\{(1+\beta_E)^{-1}\cdot\left(\triangledown{f}\times[(1+\beta_B)\cdot\bf{\dot{B}}]\right)\right\}.
\eea
And for the same reason, $\mathrm{Det}(M_b)=0$. Actually, we find that generally
$\mathrm{Det}(M_b)\propto\mathrm{Det}(M_e)$,
hence the two equations, $\mathrm{Det}(M_b)=0$ and
$\mathrm{Det}(M_e)=0$, give the same differential equations of
$f(x,y,z)$.
The former proportionality (in some cases, it even becomes an
equality) holds as expected, as the differential equation of $f$
should be unique.
This will be shown in detail in Appendix C. 
In the presence of LIV, $\mathrm{Det}(M_e)=0$ actually gives the eikonal equation
implicitly. We will show it explicitly in a simple
case in the following section.


\section{A case study and discussions\label{sec:4}}

In this section, we give explicitly the eikonal equation for a
simple case, together with more detailed discussions. For this
purpose, Eqs.~(\ref{MaE}) and (\ref{MaB}) are our starting points.

First we try to extract the tensors/matrices $M_e$ and $M_b$ from
Eqs.~(\ref{MaE}) and (\ref{MaB}) respectively. By rewriting
Eq.~(\ref{MaE}) in component form, \bea\label{te}
(M_e)^{ij}\bf{\dot{E}}^j=(1+\beta_E)^{ij}\bf{\dot{E}}^j+\epsilon^{ijk}f_j(1+\beta_B)^{kl}\epsilon^{lmn}f_m\bf{\dot{E}}^n,
\eea
where $f_i\equiv {\pt{f}}/{\pt{x^i}}$, one can read
\bea\label{mate}
(M_e)^{ij}=[1-(\triangledown{f})^2]\delta^{ij}+f_if_j+(\beta_E)^{ij}-\epsilon^{ink}\epsilon^{jml}f_nf_m(\beta_B)^{kl}.
\eea
Similarly, from Eq.~(\ref{MaB}), one can obtain 
\bea\label{matb}
(M_b)^{ij}=\delta^{ij}-\epsilon^{ink}\epsilon^{jml}f_nf_mW^{kl}-\epsilon^{isk}\epsilon^{nml}f_sf_mW^{kl}(\beta_B)^{nj},
\eea
where $W^{ij}=[(1+\beta_E)^{-1}]^{ij}$. Our task now is to
calculate the determinants of $M_e$ and $M_b$, then one can obtain the eikonal equation satisfied by $f(x,y,z)$.

As {\bf E}, {\bf B} fields are simply the 3+1 decomposition of the
electromagnetic field strength $F^{\mu\nu}$, and our working
hypothesis relies only on the analysis of causality,
we can reasonably expect that the eikonal equations obtained from
the equations of
{\bf E} 
and {\bf B} 
should be the same.

The explicit expressions of $M_e$ and $M_b$ in the matrix form are
rather tedious, not to mention the calculation of their
determinants. As an illustration, we need not to consider the full
expression with all 10 parameters non-zero. We place the discussions of the
full expression in Appendix C, where we show that the eikonal
equations obtained from Eq.~(\ref{mate}) and Eq.~(\ref{matb}) are
indeed the same. Instead, we assume here that only
$(\beta_B)^{12}=(\beta_B)^{21}=\sigma\neq0$ to simplify our discussions on 
$M_e$ and $M_b$. By virtue of this simplification, we can
subsequently get and solve the eikonal equation explicitly.

However, here we derive it by restarting derivations from the
simplified field equations,
\bea\label{SLVME}
\triangledown\cdot\bf{E}=0,\quad  \epsilon_{ijk}\partial_j\bf{B}^k+\epsilon_{ijk}\partial_j({\beta_B})^{kl}\bf{B}^l-\bf{\dot{E}}^i=0,\\
\label{SLVMB} \triangledown\cdot\bf{B}=0,\quad
~~~~~~~~~~~~~~~~~~~~~~\epsilon_{ijk}\partial_j\bf{E}^k+\bf{\dot{B}}^i=0,
\eea
where $({\beta_B})^{kl}=\sigma(\delta^k_1\delta^l_2+\delta^k_2\delta^l_1)$.
Now the equations corresponding to Eqs.~(\ref{Bligh}) and (\ref{Eligh})
are
\bea\label{SigmaB}&&\bf{\dot{B}}-(\triangledown{f})^2[(1+{\beta_B})\cdot\bf{\dot{B}}]+\triangledown{f}[\triangledown{f}\cdot({\beta_B}\cdot\bf{\dot{B}})]=\triangledown{f}\times\{\triangledown\times[(1+{\beta_B})\cdot\bf{B}_0]\}
-\triangledown\times\bf{E}_0-\triangledown{f}(\triangledown\cdot\bf{B}_0)
,\\\label{SigmaE} &&
[1-(\triangledown{f})^2]\bf{\dot{E}}+\triangledown{f}\times[{\beta_B}\cdot(\triangledown{f}\times\bf{\dot{E}})]=\triangledown{f}\times[(1+{\beta_B})\cdot(\triangledown{f}\times\bf{E}_0)]+
\triangledown\times[(1+{\beta_B})\cdot\bf{B}_0]-(\triangledown\cdot\bf{E}_0)\triangledown{f}.
\eea
From (\ref{SigmaE}) we obtain \bea (M_e)=[1-(\triangledown{f})^2]\left(
                                                   \begin{array}{ccc}
                                                    1 & 0 & 0 \\
                                                    0 & 1 & 0 \\
                                                    0 & 0 & 1 \\
                                                   \end{array}
                                                 \right)
+\sigma\left(
         \begin{array}{ccc}
           0       & f_3^2 & -f_2f_3 \\
           f_3^2   & 0     & -f_1f_3  \\
           -f_2f_3 & -f_1f_3      & 0 \\
         \end{array}
       \right).
\eea
By direct calculation, we have \bea\label{sdee}
\mathrm{Det}(M_e)=[1-(\triangledown{f})^2]\left\{[1-(\triangledown{f})^2]^2+2f_1f_2\sigma[1-(\triangledown{f})^2]-f_3^2\sigma^2(\triangledown{f})^2\right\}.
\eea
Similarly for the {\bf B} field,  from (\ref{SigmaB}) we have
\bea (M_b)=[1-(\triangledown{f})^2]\left(
                                                   \begin{array}{ccc}
                                                    1 & 0 & 0 \\
                                                    0 & 1 & 0 \\
                                                    0 & 0 & 1 \\
                                                   \end{array}
                                                 \right)
+\sigma\left(
         \begin{array}{ccc}
           f_1f_2       & f_1^2-(\triangledown{f})^2 & 0 \\
           f_2^2-(\triangledown{f})^2   & f_1f_2     & 0  \\
           f_2f_3                    & f_2f_3        & 0 \\
         \end{array}
       \right),
\eea and \bea\label{sdeb}
\mathrm{Det}(M_b)=[1-(\triangledown{f})^2]\{[1-(\triangledown{f})^2+\sigma
f_1f_2]^2-(f_1^2+f_3^2)(f_2^2+f_3^2)\sigma^2\}. \eea

One can easily verify that $\mathrm{Det}(M_e)=\mathrm{Det}(M_b)$.
Generally, the equation $\mathrm{Det}(M_e)=0$ has three solutions.
One is the conventional $(\triangledown{f})^2=1$, and if written in
terms of $w(x,y,z;t)$, it is Eq.~(\ref{LILC}). We will see below
that this corresponds to the conventional dispersion relation
$p^2=0$, which accompanies modified Lorentz-violating dispersion
relations in many models, see {\it e.g.}, Ref.~\cite{mine}. The
other two solutions, \bea\label{simls} 1-(\triangledown{f})^2+\sigma
f_1f_2\pm\sigma\sqrt{(f_1^2+f_3^2)(f_2^2+f_3^2)}=0, \eea manifest
LIV explicitly, which is characterized by the 
non-zero LIV parameter $\sigma$. Treating space and time on the same footing, {\it
i.e.}, rewriting Eq.~(\ref{simls}) in terms of function
$w(t,x,y,z)$, we can get \bea\label{simle}
(\frac{\pt{w}}{\pt{t}})^2-(\triangledown{w})^2+\sigma(w_1w_2\pm\sqrt{(w_1^2+w_3^2)(w_2^2+w_3^2)})=0,
\eea where $w_i\equiv {\pt{w}}/{\pt{x^i}}$.
When $\sigma=0$, Eq.~(\ref{simle}) reduces to the Lorentz invariant case, {\it i.e.},
Eq.~(\ref{LILC}).

In order to solve the eikonal equation, we use the Lorentz invariant
case (\ref{LILC}) as an illustration. 
Choosing the positive sign for convenience, we obtain a first
order partial differential equation, \bea\label{LIdr}
\frac{\pt{w}}{\pt{t}}+\sqrt{(\triangledown{w})^2}=0. \eea
Then from the well-known analogy between Hamilton-Jacobi equation and
geometric optics~\cite{Fock}, one can identify $w(x,y,z;t)$ as the
Hamilton action $S(q,P;t)$,
$\sqrt{(\triangledown{w})^2}=\sqrt{\sum_{i=1}^3w_i^2}$ as the
Hamiltonian $H(q,\frac{\pt{S}}{\pt{q}};t)$, and $w_i$ as the
momentum $P_i={\pt{S}}/{\pt{q^i}}$. So the question is transformed
into solving the Hamilton equation, \bea \dot{x}^i=\{x^i,
H\}=\frac{\pt{\sqrt{\sum_{i=1}^3w_i^2}}}{\pt{w_i}}=\frac{w_i}{\sqrt{\sum_{i=1}^3w_i^2}}.
\eea
We note that $\sum_{i=1}^3(\dot{x}^i)^2=1$, which can be
written in an appropriate form, $d\tau^2=dt^2-(dx^i)^2=0$,
characterizing the null geodesic of photons. If Eq.
(\ref{LIdr}) is solved in the momentum space,  we get the well-known
dispersion relation $p^2=0$ for massless particles.

Next we 
use the same analogy to solve Eq.~(\ref{simle}). In the presence
of LIV, the formal Hamilton-Jacobi equation now reads
\bea\label{HJL}&&
\frac{\pt{w}}{\pt{t}}+\sqrt{(\triangledown{w})^2-\sigma(w_1w_2\pm\sqrt{(w_1^2+w_3^2)(w_2^2+w_3^2)})}=0,
\eea  with \bea\label{Hamilton}&&
H(w_i)=\sqrt{(\triangledown{w})^2-\sigma(w_1w_2\pm\sqrt{(w_1^2+w_3^2)(w_2^2+w_3^2)})}.
\eea
Since $H$ does not contain the canonical variables which
conjugate to momentum $w_i$, it automatically implies that the
momentum $k_i=w_i$ is conserved. From the absence of explicit
dependence on time, we have ${\pt{S}}/{\pt{t}}=
{\pt{w}}/{\pt{t}}=-E=-k_0$. Then based on the above observations, we can
quickly read the dispersion relation
\bea\label{bire}
{k^0}^2=\vec{k}^2-\sigma(k^1k^2\pm\sqrt{({k^1}^2+{k^3}^2)({k^2}^2+{k^3}^2)}),
\eea
which can also be confirmed by solving field equation (\ref{eom}) (for details, see Appendix D). 

Note that the ``$\pm$'' signs in Eq.~(\ref{bire}) imply that the
photon dispersion relation depends on polarization, hence can lead
to the so-called vacuum birefringence effect~\cite{SME,CFJ}, which
has already been used to place severe constraints on LIV
parameters from astrophysical observations, like CMB~\cite{CMB},
radio galaxies~\cite{galaxy} and GRB~\cite{bire,Earbi}.

Back to the Hamilton-Jacobi equation (\ref{HJL}), by using the
observer rotational invariance, we can rotate the observer frame to
a particular one with $w_1=w_2=\rho$. 
This simplifies Eq.~(\ref{HJL}) to
\bea\label{case}
\frac{\pt{w}}{\pt{t}}+\sqrt{(\triangledown{w})^2-\sigma(\rho^2\pm(\rho^2+w_3^2))}=0.
\eea

For the positive sign in (\ref{case}), we can solve \bea\label{WFV1}
\dot{x}^1=\dot{x}^2=\hf\frac{\pt{H}}{\pt\rho}=\frac{\rho\sqrt{1-\sigma}}{\sqrt{2\rho^2+w_3^2}},
\quad
\dot{x}^3=\frac{\pt{H}}{\pt{w_3}}=\frac{w_3\sqrt{1-\sigma}}{\sqrt{2\rho^2+w_3^2}},
\eea where the factor $1/2$ comes from the symmetry $w_1=w_2=\rho$.
We can read the group velocity from Eq.~(\ref{WFV1}) as
$v_g=\sqrt{\sum_{i=1}^3{\dot{x}_i}^2}=\sqrt{1-\sigma}$, or
equivalently $d\tau^2=(1-\sigma)dt^2-d\vec{x}^2=0$. Accidentally,
the phase velocity equals the group velocity, {\it i.e.},
$v_p={H}/{|\triangledown{w}|}=\sqrt{1-\sigma}=v_g$.

For the negative sign, we have \bea\label{WFV2}
\dot{x}^1=\dot{x}^2=\hf\frac{\pt{H}}{\pt\rho}=\frac{\rho}{\sqrt{2\rho^2+(1+\sigma)w_3^2}},
\quad
\dot{x}^3=\frac{\pt{H}}{\pt{w_3}}=\frac{w_3(1+\sigma)}{\sqrt{2\rho^2+(1+\sigma)w_3^2}}.
\eea
Thus 
$d\tau^2=dt^2-[{dx_1}^2+{dx_2}^2+{dx_3}^2/(1+\sigma)]=0$, and the
group velocity reads
$$v_g=\sqrt{\frac{2\rho^2+(1+\sigma)^2w_3^2}{2\rho^2+(1+\sigma)w_3^2}}
=\sqrt{1+\frac{\sigma(1+\sigma)w_3^2}{2\rho^2+(1+\sigma)w_3^2}}\backsimeq1+
\frac{\sigma{w_3^2}}{2|\triangledown{w}|^2},$$
where in the last step we have approximated it to
the first order of $\sigma$. Similarly, the phase velocity is
$$v_p=\frac{H}{|\triangledown{w}|}=\sqrt{1
+\frac{\sigma{w_3^2}}{2\rho^2+w_3^2}}\backsimeq1+
\frac{\sigma{w_3^2}}{2|\triangledown{w}|^2}.$$ We see that only in
the first order approximation of $\sigma$, $v_p=v_g$, while
considering higher orders, $v_p\neq{v}_g$, contrary to the case of
positive sign. So in general $v_p\neq{v_g}$ in the presence of LIV, and
this is easy to understand. As the presence of the background tensor
$k_F$ breaks the Lorentz invariant vacuum, the vacuum is no longer
an isotropic and dispersion free medium. In this case, just like the
behavior of the electromagnetic wave in the conventional dispersive
medium, generally $v_p$ depends on the photon momentum $k$, {\it
i.e.}, ${\pt{v_p}}/{\pt{k}}\neq0$. Consequently,
$v_g={\pt{k^0}}/{\pt{k}}=v_p+k ~{\pt{v_p}}/{\pt{k}}\neq{v_p}$, where
$k=|\vec{k}|$.

In this case, it is interesting to define an effective refractive
index from the comparison of Eq.~(\ref{simls}) with
Eq.~(\ref{conv}),
\bea\label{refi}
n_{\mathrm{eff}}^2=1+\sigma(f_1f_2\pm\sqrt{(f_1^2+f_3^2)(f_2^2+f_3^2)}).
\eea
Similarly, it is instructive to define an effective mass
\bea\label{effm}
m_{\mathrm{eff}}^2=-\sigma(k^1k^2\pm\sqrt{({k^1}^2+{k^3}^2)({k^2}^2+{k^3}^2)}),
\eea
by comparing Eq.~(\ref{bire}) with that of the conventional
massive particles.

Then from Eqs.~(\ref{refi}) and (\ref{effm}), we see that,
for $\sigma>0$, the ``$+$'' sign gives 
$n_{\mathrm{eff}}^2>1$ and $m_{\mathrm{eff}}^2<0$. The analogy with
optical media tells us that $v_p<1$, which is confirmed in our
special case where $v_p=\sqrt{1-\sigma}$. However, the analogy with
massive particles by means of introducing an effective mass squared
$m_{\mathrm{eff}}^2$ breaks down. In the conventional case, when we
have a negative mass squared, it means that we have expanded the
theory at an unappropriate point, a point which is not a 
true vacuum. When we quantize the theory around this unstable
vacuum, we get
a tachyon with $v>1$ and hence violate causality. But this 
analogy breaks down here, as the underlying physical mechanism is
completely different from that of the false vacuum in a Lorentz
invariant theory. For the ``$-$'' sign, we have
$n_{\mathrm{eff}}^2<1$ and $m_{\mathrm{eff}}^2>0$. From the analogy
with optical media, we expect $v>1$, while naive analogy with
conventional massive particles indicates $v<1$. We see from the
special case above that
$v_p=v_g\backsimeq1+{\sigma{w_3^2}}{|\triangledown{w}|^{-2}}/2>1$,
which is
again consistent with the optical media analogy. 
For $\sigma<0$, one can obtain similar results. Thus we conclude
that the analogy between LIV vacua with optical medium is more
reasonable and helpful.

In fact, our derivation of the eikonal equation is inspired by and
starts from the analogy between the electromagnetic wave propagating
in LIV vacuo and in anisotropic media. As the
definition $n^2_{\mathrm{eff}}$ is along the same line of the
analogy, it is natural to expect that the qualitative results (of
velocity) are consistent with direct calculations. However, this is
not true for the analogy with conventional massive particles. The reason
is that here the effective mass squared is no longer a free
parameter in Lagrangian; instead, it is momentum-dependent. More
importantly, now Lorentz invariance is broken, thus it is not
necessary for the maximum attainable velocity of the effective
massive particle to be $c=1$.

Moreover, we see that, just as what happened in the conventional
dispersive medium~\cite{superl}, in our case, the definition of
light velocity is more involved than that in the conventional
Lorentz invariant vacuo. Perhaps it is more complicated, for now it
is the vacuum itself becoming anisotropic. Though in the dispersive
medium, there already exist cases with $v_p>c$ or
$v_g>c$~\cite{superl}, they do not conflict with the
requirement of causality. Since in practice, we can only measure
directly the velocity of light signals, denoted as $v_s$. The only
difficulty in that case is a proper definition of signal velocity.
Once we have an appropriate definition of $v_s$, it can be shown
that $v_s<1$~\cite{brill}.
Thus no causal problem arises there. Since in that case, our starting point, the
wave equation, does indeed obey Lorentz invariance. Only the medium
in which light propagates, is no longer the Lorentz invariant
vacuum, but crystal or nontrivial QED vacuum. The medium then
singles out a preferred direction, which leads to the so-called soft
breaking of Lorentz invariance~\cite{FTL}, with a superluminous
phase/group velocity.

However, this is not the situation here. Now it is the basic wave
equation, Maxwell equation, which is no longer Lorentz invariant.
Hence there might be more difficulties in the proper definition of
signal velocity and it may cause causal problems. For the simple
case above, we can define $v_s$ as the wavefront velocity $v_f$.
Then since our calculation deals directly with electromagnetic
wavefront, we find $v_g=v_f$. Alternatively, we can also define
energy transport velocity $\vec{v}_e^i=\Theta^{i0}/\Theta^{00}$ as
the signal velocity. Then by choosing the appropriate ansatz
$A_1(t,\vec{r})_\pm=W_\pm
\mathrm{cos}[w(t-\vec{s}\cdot\vec{r}/v_{p\pm})]$ (where ``$\pm$''
sign refers to the two modes in Eq.~(\ref{bire}), respectively) and the Coulomb
gauge $\triangledown\cdot\vec{A}=0$, we can prove
$|\vec{v}_e|=v_g$
for both ``$+$'' and ``$-$'' 
modes in Eq.~(\ref{bire}) (for details of this issue, see
Appendix D). Thus for the simple cases corresponding to
Eqs.~(\ref{WFV1}) and (\ref{WFV2}), and for either sign of $\sigma$, we see
that there always exists one mode with
$v_s>1$, which indeed violates causality. However, note that all the
treatments above are purely classical. Since LIV is believed to be a
quantum gravitational phenomenon, the best way to treat the corresponding causality problem
is to use quantum field theory. Hence it is more appropriate to discuss the
microcausality from the
calculation of 2-point correlation functions of observables at
space-like separations. However, this is beyond the scope of the
present paper.

\section{Conclusion}

Inspired by the analogy between the behavior of LIV electrodynamics
and electrodynamics in macroscopic anisotropic medium, we derived the eikonal
equation for Lorentz non-invariant vacuum from the modified Maxwell
equations of SME. The results obtained from the equations of both
{\bf E} and {\bf B} fields are the same, and the general case is
given in Appendix C. This implies that the derivation is
self-consistent. Then we use the well-known analogy between geometric
optics and the Hamilton-Jacobi equation, and find that the solution
of the eikonal equation (\ref{simls}) in the momentum space, {\it
i.e.}, Eq.~(\ref{bire}), turns out to be the modified dispersion relation of
photons, which is the same as those obtained from other approaches.
This fact confirms the consistency of our approach further.

With the definition of an effective refractive index
$n_{\mathrm{eff}}$, we find that 
the dependence of
the velocity on the LIV parameter 
is consistent with 
the velocity dependence on the refractive index $n$ of ordinary
refractive medium. As our calculation deals directly with wavefront,
the results show that the front velocity equals the group
velocity for each LIV mode. From the analogy of the LIV vacuum with
macroscopic anisotropic medium, we find that the presence of LIV
makes the proper definition of signal velocity more complicated,
especially when considering the existence of multi-definition of
velocity. Therefore, the treatment of causality becomes much subtle.
However, if we define the signal
velocity as the wavefront velocity, {\it i.e.}, $v_s=v_f$, 
naive analysis shows that, for either sign of $\sigma$,
causality is violated classically at least for one mode. 
Of course, to see whether or not tiny LIV might threaten microcausality, one
should turn to the more trustable and rigorous treatment of
quantum field theory to calculate the correlation
functions at space-like separations. Actually, this was done for the
Chern-Simons-like term~\cite{CSCU} and it has caused extensive
debates in the literature 
(see {\it e.g.}, Ref.~\cite{CSCU}). Complete treatment of
microcausality was also achieved for the massive fermion sector of
SME~\cite{SME,spont}, and it was shown that no crucial problems
arise for a spontaneous Lorentz invariance breaking theory. If
Lorentz violation happens intrinsically at high energies rather than
spontaneously, as suggested in the so-called weighted dimensional
model, Anselmi pointed out that it is Bogoliubov's criterion of
causality, rather than the value of correlation functions, that
makes sense~\cite{DA}.

As a by-product of this work, we find that the asymmetric
energy-momentum tensor $\Theta^{\mu\nu}$ causes no inconsistency in
principle. By assuming rotational invariance, the conservation of
Noether charge, {\it i.e.}, $d\Sigma_{ij}/dt=0$, requires that the
spatial components of $\Theta^{\mu\nu}$ must be symmetric. Through
inspection of the corresponding LIV parameters in the
presence of rotational symmetry, we find that indeed 
$\Theta^{ij}=\Theta^{ji}$, though this is valid only in a particular reference frames,
as already mentioned. While from the definitions of
$\Theta^{\mu\nu}$ and the velocity of energy transport, in our simple case,
explicit calculation shows that the group velocity also equals the energy
transport velocity for each mode. Together with
$v_g=v_f$, it makes the definition of signal velocity as wavefront
velocity more reliable in the classical discussion of causality. The
last point we would like to stress is that the derivation of the
eikonal equation is not restricted to the simple model here. In fact,
at least for gauge invariant Maxwell equations without {\bf E}, {\bf
B} mixing terms, it should work as well. Thus the method might be
helpful in the causality and phenomenological analysis of
various LIV extensions of Maxwell equations.

\section*{Acknowledgments}

We thank Liang Zhang, Zhi-bo Xu for helpful discussions. This work
is partially supported by National Natural Science Foundation of
China (11005018, 10721063, 10975003, 11035003), by the Key Grant Project of Chinese
Ministry of Education (No.~305001), and by the Research Fund for the
Doctoral Program of Higher Education (China). It is also supported
by National Fund for Fostering Talents of Basic Science (Nos.
J0630311, J0730316) and Hui-Chun Chin and Tsung-Dao Lee Chinese
Undergraduate Research Endowment (Chun-Tsung Endowment) at Peking
University.


\section*{Appendix A}
This appendix gives the proof that, even for a generic asymmetric energy-momentum tensor,
the requirement of rotational invariance implies the symmetric relation $\Theta^{ij}=\Theta^{ji}$,
which ensures the consistency with the corresponding Noether current. Here, the
conserved charge corresponding to the Noether current is the
angular momentum.

For a Lorentz invariant theory, the conservation of generators of
Lorenz group is manifested through the Noether current,
\be\label{HLC}
M^\rho_{~\mu\nu}=\hf(\Theta^\rho_{~\nu}x_\mu-\Theta^\rho_{~\mu}x_\nu).
\ee
One can construct the corresponding generators of Lorentz group as
\bea
\Sigma_{\mu\nu}=\int{d^3x}M^0_{~\mu\nu}. \eea

Then from the current conservation,
\be\label{Nother}
\pt_\rho{M^\rho_{~\mu\nu}}=0, \ee
it is easy to show that
$d\Sigma_{\mu\nu}/dt=0$. Hence the six generators of Lorentz group
are conserved.
Meanwhile, combined with the conservation of energy-momentum
tensor, \emph{i.e.}, $\pt_\rho\Theta^{\rho\mu}=0$, the current
conservation implies that, \bea\label{SYMM}&&
0=\pt_\rho{M^{\rho\mu\nu}}=\hf[x^\mu\pt_\rho\Theta^{\rho\nu}-x^\nu\pt_\rho\Theta^{\rho\mu}+(\Theta^{\rho\nu}\delta_\rho^\mu-\Theta^{\rho\mu}\delta_\rho^\nu)]\nn
&&=\hf(\Theta^{\mu\nu}-\Theta^{\nu\mu}), \eea
 {\it i.e.}, the
energy-momentum tensor is symmetric. However, as mentioned in the
main text, generally this is not valid in the presence of LIV.

Here we note that the same reasoning above can also be applied to
the rotational symmetry. If we only break boost invariance, then the
infinitesimal symmetry transformation \bea\label{LT}
\Psi^l(x)\rightarrow{\Psi^l(x)-\frac{i}{2}w^{\mu\nu}(\mathscr{J}_{\mu\nu})^l_{~m}\Psi^m(x)}
\eea  under Lorentz invariance is replaced with \bea
\Psi^l(x)\rightarrow{\Psi^l(x)-\frac{i}{2}w^{ij}(\mathscr{J}_{ij})^l_{~m}\Psi^m(x)}
\eea under rotational invariance. It is nearly the same as
(\ref{LT}) except that $w^{\mu\nu}$ is replaced by $w^{ij}$ (where
as before, Latin indices $i$, $j$ run over the three spatial
coordinate labels, usually taken as 1, 2, 3; and Greek indices
$\mu$, $\nu$ run over the four space-time coordinate labels, 1, 2, 3,
0).

With the above observation, it is apparent that from the Noether
theorem, we can obtain the corresponding Noether current
$M^\rho_{~ij}$ and Noether charge $\Sigma_{ij}$. With the same
reasoning, $\pt_\rho{M^{\rho{ij}}}=0$ also implies
$\Theta^{ij}=\Theta^{ji}$.

The symmetrizability of $\Theta^{ij}$ is valid for a generic LIV
theory with rotational symmetry SO(3) unbroken, like the photon
Lagrangian with an anisotropic scaling in Ref.~\cite{mine}. However, we stress again that
this symmetric property and SO(3) invariance is valid only in a specific inertial reference frames,
because LIV implies the existence of a preferred direction. While
for a specific Lagrangian (\ref{rp}) without CPT odd terms, we have
\bea
\Theta^{\mu\nu}=-[F^{\mu\alpha}F^\nu_{~\alpha}+(k_F)^{\alpha\beta\mu\delta}{F^\nu_{~\delta}}F_{\alpha\beta}]-\eta^{\mu\nu}\mathcal{L},
\eea
so
\bea
\Theta^{\mu\nu}-\Theta^{\nu\mu}=-[(k_F)^{\alpha\beta\mu\delta}F^\nu_{~\delta}-(k_F)^{\alpha\beta\nu\delta}F^\mu_{~\delta}]F_{\alpha\beta}
\eea
and
\bea\label{rota}
\Theta^{[ij]}&=&\hf(\Theta^{ij}-\Theta^{ji}) \nn
&=&{\hf}F_{\alpha\beta}[(k_F)^{\alpha\beta{j}\delta}F^{i}_{~\delta}-(k_F)^{\alpha\beta{i}\delta}F^{j}_{~\delta}]\nn
&
=&\hf{F_{kl}}[((k_F)^{kljm}F^i_{~m}-(k_F)^{klim}F^j_{~m})+((k_F)^{klj0}F^i_{~0}-(k_F)^{kli0}F^j_{~0})]\nn
&&+F_{k0}[((k_F)^{k0jl}F^i_{~l}-(k_F)^{k0il}F^j_{~l})+((k_F)^{k0j0}F^i_{~0}-(k_F)^{k0i0}F^j_{~0})].
\eea

Since rotational invariance requires only $\al\neq0$, from
(\ref{prde}) and (\ref{myde}), we have \bea\label{roint}&&
(k_F)^{kljm}=\frac{\al}{2}(\delta^{kj}\delta^{lm}-\delta^{km}\delta^{lj}),
\quad  (k_F)^{kli0}=0,\quad (k_F)^{k0i0}=-\frac{\al}{2}\delta^{ki}.
\eea
Substituting (\ref{roint}) into (\ref{rota}), one can easily
find that
\bea
\Theta^{[ij]}=\frac{\al}{2}[(F_{im}F^j_{~m}-F_{jm}F^i_{~m})+(F_{i0}F^j_{~0}-F_{j0}F^i_{~0})]=0,
\eea
which is consistent with the requirement of rotational
invariance.
The same property of $\Theta^{ij}$ can also be checked for more
general theories with rotational invariance, {\it e.g.}, the
Horava-Lifschitz theory.
\section*{Appendix B}
This appendix shows how to get (\ref{LILC}) from (\ref{conv}).

From (\ref{exlic}), we have
\bea\label{app1}
dt=f_idx^i~~~~\Rightarrow\frac{\pt{t}}{\pt{x^i}}=f_i.
\eea
While from (\ref{imlic}), we get
\bea\label{app2} 0\equiv
dw=w_idx^i+w_tdt~~~\Rightarrow~~~\frac{\pt{t}}{\pt{x^i}}=-\frac{w_i}{w_t},
\eea
where $w_i\equiv {\pt{w}}/{\pt{x^i}}$ and $w_t\equiv
{\pt{w}}/{\pt{t}}$. Then from (\ref{app1}) and (\ref{app2}), one can
easily obtain $ \triangledown_if=-{w_i}/{w_t}$. So we have
\bea
1=(\triangledown{f})^2=\sum_{i=1}^3(w_i/w_t)^2=(\triangledown{w})^2/(w_t)^2.
\eea

\section*{Appendix C}
In this appendix we show that the eikonal equations we got from
(\ref{mate}) and (\ref{matb}) are the same, and hence are consistent with
the general arguments in the main text: as the derivation roots in
the classical causality analysis, and \textbf{E}, \textbf{B} fields
are components of the $3+1$ decomposition of $F^{\mu\nu}$, hence being
connected with each other through the Maxwell equations, the eikonal
equations derived from the equations of \textbf{E} and \textbf{B} must
be the same.

First, let us focus on the case $(\beta_E)^{ij}=0$. The consistent
check of this simpler case will be a little different from the case
$(\beta_E)^{ij}\neq0$ . For $(\beta_E)^{ij}=0$, we denote the
corresponding matrices of (\ref{mate}) and (\ref{matb}) as $M_{eB}$
and $M_{bB}$ respectively, {\it i.e.}, \bea\label{MEB}  &&
(M_{eB})^{ij}=[1-(\triangledown{f})^2]\delta^{ij}-\epsilon^{ink}\epsilon^{jml}f_nf_m(\beta_B)^{kl},
\\\label{MBB}&&
(M_{bB})^{ij}=[1-(\triangledown{f})^2]\delta^{ij}-(\triangledown{f})^2(\beta_B)^{ij}-f_if_k(\beta_B)^{kj}.
\eea 
If we assign \bea \beta_B=\left(
          \begin{array}{ccc}
            b_4 & b_1 & b_2      \\
            b_1 & b_5 & b_3      \\
            b_2 & b_3 & -(b_4+b_5)\\
          \end{array}
        \right),
\eea the matrices above take the explicit matrix forms as below
{\tiny \bea (M_{eB}) =\left(
           \begin{array}{ccc}
               [1-(\triangledown{f})^2]+f_2^2(b_4+b_5)-f_3^2b_5+2f_2f_3b_3 & f_3^2b_1-f_2f_1(b_4+b_5)-f_3(b_3f_1+b_2f_2) & f_1f_3b_5+b_2f_2^2-f_2(b_3f_1+b_1f_3) \\
               f_3^2b_1-f_2f_1(b_4+b_5)-f_3(b_3f_1+b_2f_2) & [1-(\triangledown{f})^2]+f_1^2(b_4+b_5)-f_3^2b_4+2f_1f_3b_2 & f_2f_3b_4+b_3f_1^2-f_1(b_2f_2+b_1f_3) \\
               f_1f_3b_5+b_2f_2^2-f_2(b_3f_1+b_1f_3) & f_2f_3b_4+b_3f_1^2-f_1(b_2f_2+b_1f_3) & [1-(\triangledown{f})^2]-(f_2^2b_4+b_5f_1^2)+2f_2f_1b_1 \\
             \end{array}
  \right)\nonumber
\eea \bea (M_{bB})=\left(
             \begin{array}{ccc}
               [1-(1+b_4)(\triangledown{f})^2]+f_1(b_4f_1+b_1f_2+b_2f_3) & f_1(b_1f_1+b_5f_2+b_3f_3)-b_1(\triangledown{f})^2 & f_1(b_2f_1+b_3f_2-(b_4+b_5)f_3)-b_2(\triangledown{f})^2 \\
               f_2(b_4f_1+b_1f_2+b_2f_3)-b_1(\triangledown{f})^2 & [1-(1+b_5)(\triangledown{f})^2]+f_2(b_1f_1+b_5f_2+b_3f_3) & f_2(b_2f_1+b_3f_2-(b_4+b_5)f_3)-b_3(\triangledown{f})^2 \\
               f_3(b_4f_1+b_1f_2+b_2f_3)-b_2(\triangledown{f})^2 & f_3(b_1f_1+b_5f_2+b_3f_3)-b_3(\triangledown{f})^2 & [1-(1-b_4-b_5)(\triangledown{f})^2]+f_3(b_2f_1+b_3f_2-(b_4+b_5)f_3) \\
             \end{array}
           \right).\nonumber
\eea}
Note that $(M_{eB})^T=M_{eB}$, {\it i.e.}, $M_{eB}$ is a symmetric matrix, while $M_{bB}$ is not. So it is not a trivial check 
that $\mathrm{Det}[M_{eB}]=\mathrm{Det}[M_{bB}]$ by direct
calculation. Imposing the requirements that their determinants be equal
to zero, one can obtain the same equations, as expected. Since the
results are tedious, we do not present them here.
On the other hand, if we allow only $(\beta_B)^{21}=(\beta_B)^{12}\neq0$, one can
easily get back to (\ref{sdee}) and (\ref{sdeb}). This can be
regarded as another consistent check.

For the case $(\beta_E)^{ij}\neq0$, as the calculation
involves the inverse of $(1+\beta_E)^{ij}$, {\it i.e.},
$W^{ij}=[(1+\beta_E)^{-1}]^{ij}$, the results will be more tedious
than the previous ones. So we also do not present the details here. We just
make two remarks.
\begin{enumerate}
\renewcommand{\labelenumi}{(\theenumi)}
\item
As the calculation of (\ref{matb}) involves the matrix $W^{ij}$,
while that of (\ref{mate}) does not, the determinants of
(\ref{mate}) and (\ref{matb}) are indeed not the same, contrary to
the previous case. However, note that the requirement of the same
eikonal equations is derived from the vanishing of the determinants
of matrices (\ref{mate}) and (\ref{matb}). This requires only
that the determinants of these two matrices are proportional to each
other, not necessarily being the same. In fact, 
one can show that they differ only by a constant, {\it i.e.},
\bea\label{comp}
\mathrm{Det}[M_e]=\mathrm{Det}[M_b]\cdot\mathrm{Det}[1+\beta_E].
\eea So as long as $\mathrm{Det}[1+\beta_E]\neq0$, the differential
equations obtained from $\mathrm{Det}[M_e]=0$ and
$\mathrm{Det}[M_b]=0$ must be the same.
\item
By taking $\beta_E=0$, the solution reduces to those obtained from
(\ref{MEB}) or (\ref{MBB}), except for a missing factor
$1-(\triangledown{f})^2$.  This can be traced back to solving field equations,
and our calculation is implicitly
equivalent to that procedure.  
In solving Maxwell equations, in order to obtain the independent equations of the two
physical degrees of freedom, we need to choose a gauge, {\it e.g.},
Coulomb gauge, to eliminate the gauge degrees of freedom (see
Appendix D). In this process, we eliminate one polarization and
leave a constraint, which in turn gives an identity. Eventually, we
are left with only two independent equations, corresponding to the
two physical polarizations. The eliminated identity in this process
then corresponds to the lacking factor in our method. In our
derivation, it is the $f_if_j$ factor in (\ref{mate}) (or the
$-\epsilon^{ink}\epsilon^{jml}f_nf_mW^{kl}$ factor in (\ref{matb}))
playing the role of gauge constraints to remove
$1-(\triangledown{f})^2$ in (\ref{comp}).
\end{enumerate}

In conclusion, we see that, as
expected, the calculation from either (\ref{mate})
or (\ref{matb}) indeed leads to the same eikonal equations, and the derivation is also consistent with solving field
equations. Thus it is natural to get from the eikonal equations two
independent dispersion relations, corresponding to two physical
polarizations, see {\it e.g.}, (\ref{bire}).
\section*{Appendix D}
In this appendix we give an explicit calculation of the
energy-momentum flow velocity $\vec{v}_e^i=\Theta^{i0}/\Theta^{00}$.

Using the ansatz
$A^\mu(x)=\epsilon^\mu(p)\mathrm{exp}[-i(p^0t-\vec{p}\cdot\vec{x})]$
and choosing the Coulomb gauge $\triangledown\cdot\vec{A}=0$, we get
$\phi=0$ as a special solution from $\triangledown\cdot\vec{E}=0$ ({\it i.e.}, the first
equation of (\ref{SLVME})).
Then by substituting Eq.~(\ref{FEB}) into Eqs.~(\ref{SLVME}) and (\ref{SLVMB}), we find 
that (\ref{SLVMB}) is satisfied automatically. Then from the left
equation of (\ref{SLVME}), we get
\bea\label{inmed}  &&
-p^2A_1+\sigma{p_3}(p_2A_3-p_3A_2)=0, \nn  &&
-p^2A_2+\sigma{p_3}(p_1A_3-p_3A_1)=0, \nn  &&
-(p^2+2\sigma{p_1p_2})A_3+\sigma(p_1A_2+p_2A_1)p_3=0.
\eea
By imposing the Coulomb gauge $p_iA_i=0$, the equations above can be
reduced to
\bea\label{last} \left(
  \begin{array}{cc}
    p^2+\sigma{p_1p_2} & \sigma(p_3^2+p_2^2) \\
    \sigma(p_3^2+p_1^2) & p^2+\sigma{p_1p_2} \\
  \end{array}
\right)\{\begin{array}{c}
               A_1 \\
               A_2
             \end{array}\}=0.
\eea
Note that, in this appendix, we do not distinguish the upper and lower indices,
{\it e.g.}, $A_1=A^1$. From the existence of
non-zero solutions of (\ref{last}), one can easily get the dispersion
relation (\ref{bire}). To calculate the energy-momentum flux
velocity, we use the real component of our previous ansatz instead,
{\it i.e.}, $A_1(t,\vec{r})_\pm=W_\pm
\mathrm{cos}[w(t-\vec{s}\cdot\vec{r}/v_{p\pm})]$, where $\vec{s}$
denotes the unit vector pointing to the direction of propagation
and $v_{p\pm}$ denote the two independent phase velocities of the
two modes in (\ref{bire}). With the help of (\ref{last}), one can
express $\vec{A}$ in terms of $A_1$, {\it i.e.},
\bea\label{component}
A_2=-\frac{p^2+\sigma{p_1p_2}}{\sigma(p_3^2+p_2^2)}A_1,\quad
A_3=-\frac{(p_2A_2+p_1A_1)}{p_3}.
\eea
Substituting these back to
(\ref{FEB}), we can obtain the explicit forms of \textbf{E} and
\textbf{B}, where we have assigned
\bea
\vec{s}=(\frac{1}{\sqrt{2}}\cos[\theta],~\frac{1}{\sqrt{2}}\cos[\theta],~\sin[\theta]).
\eea
Then by substituting the explicit forms of \textbf{E} and
\textbf{B} into (\ref{Poyn}) and (\ref{EnDe}), we find that indeed
$|\vec{v}_{e\pm}^i|=v_{g\pm}$ for each mode.
%
%

\end{document}